\documentclass[aps,pre,floatfix,nofootinbib]{revtex4}
\usepackage{graphicx} 
\usepackage[sort&compress]{natbib}
\usepackage{graphicx}
\usepackage{amsmath}\usepackage{amssymb}
\usepackage{subfigure}
\newcommand{\BEQ}{\begin{equation}}
\newcommand{\EEQ}{\end{equation}}
\newcommand{\BEA}{\begin{eqnarray}}
\newcommand{\EEA}{\end{eqnarray}}
\newcommand{\comment}[1]{}
\newcommand{\Q}{{\cal Q}}
\renewcommand{\S}{{S}}
\renewcommand{\L}{{L}}
 
\begin{document}
\draft

\title{Emergence of Leadership in Communication }

\author{Armen E. Allahverdyan$^{1 *)}
$ and Aram Galstyan$^{2)}$}

\address{$^{1)}$Yerevan Physics Institute, Alikhanian Brothers Street
2, Yerevan 375036, Armenia,\\
$^{2)}$  USC Information Sciences Institute, 4676 Admiralty Way, 
Marina del Rey, CA 90292, USA \\
{\rm * Corresponding author: 
armen.allahverdyan@gmail.com}
}

\date{\today}

\begin{abstract} 

  We study a neuro-inspired model that mimics a discussion (or
  information dissemination) process in a network of agents. During
  their interaction, agents redistribute activity and network weights,
  resulting in emergence of leader(s). The model is able to reproduce
  the basic scenarios of leadership known in nature and society:
  laissez-faire (irregular activity, weak leadership, sizable
  inter-follower interaction, autonomous sub-leaders); participative
  or democratic (strong leadership, but with feedback from followers);
  and autocratic (no feedback, one-way influence). Several pertinent
  aspects of these scenarios are found as well---e.g., hidden
  leadership (a hidden clique of agents driving the official
  autocratic leader), and successive leadership (two leaders influence
  followers by turns). We study how these scenarios emerge from
  inter-agent dynamics and how they depend on behavior rules of
  agents---in particular, on their inertia against state changes.

\end{abstract}






\maketitle

\section*{\large Introduction}

The notion of opinion leaders has become paradigmatic in social
sciences \cite{encyclo,winkler}.  Identifying leaders can be important
for applications such as viral marketing, accelerating (or blocking)
the adoption of innovations, {\it etc}. Communication research
postulates that informational influence in groups often happens via a
two-step process, where information first flows from news media to
opinion leaders, and then is spread further to followers
\cite{lazar}. This theory is believed to adequately account for
consumer behavior \cite{beckman} and has been refined in several ways
\cite{trol,burt}.

A significant research has focused on identifying traits and
characteristics of leaders. For instance, it has been observed that
opinion leaders are {\it not} necessarily more educated than
followers, but they typically have higher income \cite{corey}.
However, it was understood that no single trait (or even a cluster of
traits) can explain the emergence of leadership \cite{stogdill}. It
is believed that leaders can be imposed externally, or emerge within
the group \cite{valente}|possibly out of purely random reasons
\cite{madow}|together with agents who are not opinion leaders, but are
important for ensuring that the leaders function \cite{weimann}.


Social psychology developed several qualitative theories on how
leaders perform in groups \cite{hogg}.  The contingency
theory|developed in opposition to ``leadership trait'' approaches|
focuses on the nature of interactions between the leader and followers
\cite{lewin,hogg,fi}. According to this theory, the effectiveness of a
particular type of leadership is contingent on the favorability of the
situation to that type.  Based on earlier research \cite{lewin}, the
contingency theory identified several major types of leaders
\cite{fi}:

-- Laissez-faire leadership is characterized by a relatively weak
guidance of autonomous followers. It occurs in systems such as
scientific collectives.

-- Participative (democratic) leaders do influence their followers 
strongly, but encourage and accept feedback from them.

-- No feedback (from the majority of followers) is accepted by
autocratic leaders. This type of leadership is typically present in
military, businesses and governance systems.

Given the above typology, which is well-confirmed by everyday
experience, it is not completely clear which features imply specific
leadership types \cite{encyclo,winkler}. Our main purpose is to study
this problem via a mathematical model.


The network theory developed methods for identifying opinion leaders
via the core-periphery (or star) structure of social networks
\cite{bocca,blondel,luck,freeman,everett,puck}. Such structures can
emerge from imposing on the network certain functional goals or
optimization principles \cite{sole,subra,cholvi,ardeshir}. Recent
research on social networks proposed several methods for uncovering
hidden network of influences
\cite{ha,huber,gomez_leskovec,barba1,barba2}, and identifying
core-periphery structures related to opinion leaders
\cite{gomez_leskovec}.

Several models were proposed for describing the leadership
phenomenon. Ref.~\cite{colomer} considers game-theoretic models, where
the leadership is defined via the possibility of the first move in a
game. The related notion of Stackelberg equilibrium is one of the
basic game-theoretic manifestations of leadership \cite{sta}. Models
of adaptive (inhomogeneous) networks identify emergent leaders with
well-connected nodes in a network of game-theoretic
\cite{zimmo,anghel} or resource-distribution units
\cite{guttenberg,lipo}. Other approaches to emergent leadership are
reviewed in \cite{vugt,hazy}. The leadership problem relates to
diversity modeling in social and biological systems \cite{bona,naim};
see \cite{castel} for a review.

Despite of much inter-disciplinary research devoted to the leadership
issue, we seem to lack a basic model that can reproduce in a single
set-up the major leadership scenarios, and relate them to behavior of
involved agents.

We intend to provide a formal, mathematical model for the emergence
and the type of leadership in a collective of interacting agents
modeled via neurons. Certain analogies between human agents and
neurons were noted in literature
\cite{rashevsky1,rashevsky2,rashevsky3,rashevsky4,
  klopf,shvyrkov,e-y1,e-y2,nowak,vidal,larson,arenas,armen_aram}. Both
neurons and human agents are adaptive entities that form communities,
analyze information, and can specialize for different roles and
functions in (resp.)  brain and society \cite{klopf}. Quantitative
sociology employs neuronal models for describing weak ties
\cite{weak_ties}, social impact \cite{granovetter}, and economic
activity \cite{morris,durlauf,cortes}. Neuronal models are capable to
generalize on the concept of activity cascade that is frequently
employed in social modeling \cite{watts, dodds_watts,galstyan}. The
major limitation of this concept is that the activation of each
network node occurs only once. However, there is an ample of empiric
evidence that social activity patterns are recurrent
\cite{cascades_recur,barba1,barba2}, an effect well described by neuronal
models \cite{arenas,armen_aram}.


Our model mimics a discussion process, where opinion expression by one
agent facilitates activation of other agents. Experimental studies on
the leadership emergence in discussion groups were carried out in
\cite{babble1,babble2}.  It was noted there that the leader emerges
due to it active involvement into the group dynamics, e.g. due to
active talking, while the quality of this talking is not very
important (babble effect) \cite{babble1,babble2}.

We postulate tractable rules for the agent's behavior. These rules
incorporate major factors that are relevant for the leadership,
e.g. activity, attention, initial social capital
(i.e. well-connectedness in the network), and score (credibility).
The rules depend on parameters that characterize the agent's
``conservatism'' with respect to changing its state.

The leader is naturally defined as an agent that influences other
agents strongly (i.e. stronger than those agents influence the
leader), and that actively participate in the group activity, in the
sense that blocking the leader will diminish (or at least essentially
decrease) the activity of the group.

Our main result is that the three basic leadership
scenarios|laissez-faire, participative and autocratic|emerge under
different behavioral rules. Here is a short list of certain specific
leadership features predicted by the model.

The laissez-faire leadership emerges under noisy, but score-free
dynamics. It relates to irregular activity patterns, allows autonomous
sub-leaders and sizable communication between followers. In the
participative situation possible sub-leaders are strict subordinates
of the leader. A participative leader emerges due to an initial
(possibly small) advantage in its social capital. If the inter-agents
interaction is sufficiently strong, the emergent participative leader
wins over an externally imposed candidate for leadership. The
autocratic situation is more vulnerable to an externally imposed
leader, than the participative one. There are cases, where the
``official'' autocratic leader is driven by a hidden click of other
agents. Coalitions of autocratic leaders are possible, but they are
meta-stable, and for long times reduce to just two leaders driving the
followers by turns.

\section*{\large The model}
\label{model}

\subsection*{State Dynamics}

The main ingredients of our model are listed in Table \ref{tab1}.  We
consider $N$ agents. At a given moment of discrete time $t$, each
agent can be active|give an opinion, ask question {\it etc}|or
passive. For each agent $i$ ($i=1,...,N$) we introduce a variable
$m_i(t)$ that can assume two values $0$ (passive) and $1$ (active).

Following a tradition in quantitative sociology \cite{weak_ties,
  granovetter,morris,durlauf,cortes,watts,dodds_watts,galstyan,arenas},
we model agents via thresholds elements, i.e. we postulate that each
agent $i$ has an information potential $w_i(t)\geq 0$, and $i$
activates whenever $w_i(t)$ is larger than a threshold $u_i>0$:
\begin{eqnarray}
\label{n1}
m_i(t)&=&\vartheta [\, w_i(t)-u_i\,],~~t=0,1,2,...\, ,\\
w_i(t+1)&=&(1-m_i(t)\,) \sum_{j=1}^N q_{ij}(t)m_j(t),
\label{n2}
\end{eqnarray}
where $\vartheta(x)$ is the step function: $\vartheta (x< 0)=0$,
$\vartheta(x\geq 0)=1$. The factor $(1-m_i(t))$ in (\ref{n2})
nullifies the potential after activation; hence an agent cannot be
permanently active.  The influence $q_{ij}(t)m_j(t)$ of $j$ on $i$ is
non-zero provided that $j$ activates, $m_j(t)=1$. We assume that
$q_{ij}\geq 0$ and $q_{ii}=0$, i.e. connections can only facilitate
the potential generation.  Given the freedom in choosing $q_{ij}$, we
take $u_i=1$.


The continuous-time limit of (\ref{n1}, \ref{n2}) reduces to an
integrate and fire model of neuronal dynamics
\cite{peto,abbott,schuster}.

\begin{table*}
\caption{Here we list the main ingredients of the model
    with relevant notations and equations that introduce them.}
\begin{tabular}{||c|c||}
  \hline
  \hline
  {\bf Equations and references}      & {\bf Ingredients of the model}     \\
  \hline
  \hline
  $i=1,...,N$  & $N$ agents modeled as neurons in discrete time $t=1,2,...$. \\
  \hline
  Eqs~(\ref{n1}, \ref{n2}). Refs.~\cite{peto,abbott,schuster}.     
  & Firing rule for activity $m_i(t)$ of discrete agents-neurons. \\
  \hline
  Eqs~(\ref{eq:900}, \ref{eq:23}).  Ref.~\cite{malsburg}.
  & Constrained network weights $\tau_{ij}$ model attention limitation.   \\
  \hline
  Eq~(\ref{eq:35}).      & Adaptation of weights: an active agent gets
  more attention \\
  & from others. Parameter $\alpha$ accounts for inertia.     \\
  \hline
  Eqs~(\ref{eq:355}, \ref{eq:400}).    
  & An agent which gets more attention obtains larger score 
  $\sigma_i$.     \\
  & The score-weight interaction is controlled by parameter $\beta$. \\
  \hline
  Eqs~(\ref{noise}, \ref{n33}). Refs.~\cite{peto,abbott,schuster}.        
  & Behavioral noise $\phi_i(t)$ with magnitude $\eta$.     \\
  \hline
  \hline
\end{tabular}
\label{tab1}
\end{table*}

In (\ref{n2}), $q_{ij}(t)$ quantifies the influence of $j$ on $i$.
We parametrize it as
\begin{eqnarray}
  \label{eq:90}
 && q_{ij}(t)=q\,\tau_{ij}(t), \quad \tau_{ii}(t)=0, \\
 && \sum_{j=1}^N \tau_{ij}(t)=1, 
  \label{eq:900}
\end{eqnarray}
where $q$ is the maximal possible value of $q_{ij}(t)$. Now
$\tau_{ij}$ is the weight of influence. Eq~(\ref{eq:900})
reflects the fact that agents have limited attention
\cite{hu_at,v_at,kristina}. This characteristics was also noted for neurons
\cite{malsburg}, and it is achieved via introducing non-normalized
weights $\widetilde{\tau}_{ij}(t)$ \cite{malsburg}:
\begin{eqnarray}
  \label{eq:23}
  \tau_{ij}(t)= \widetilde{\tau}_{ij}(t)\left /
  \sum_{j=1}^N \widetilde{\tau}_{ij}(t)\right. .
\end{eqnarray}
Importantly, we do not pre-determine the network structure. The sum
$q{\sum}_{j=1}^N\tau_{ij}(t)m_j(t)$ in (\ref{n2}) is taken over all
the agents, and the weights change in time as
\begin{eqnarray}
  \widetilde{\tau}_{ij}(t+1) = \tau_{ij}(t)
+f\,\tau^{\alpha}_{ij}(t) \,
  {m}_{j}(t+1), \qquad \alpha>0,
  \label{eq:35}
\end{eqnarray}
where a non-active $j$ ($m_{j}(t+1)=0$) does not change
$\tau_{ij}$. In one version of the model $f=$const. Thus
$\widetilde{\tau}_{ij}(t)$ changes such that more active and more
credible agents get more attention from neighbors. In (\ref{eq:35}),
$\tau^{\alpha}_{ij}(t)$ controls the extent to which $i$ re-considers
those links that did not attract its attention previously
(conservatism): for $\alpha\not\approx 0$ the weight with
$\tau_{ij}(t)\approx 0$ is not reconsidered in the next step. A
similar structure was employed for modeling confirmation bias
\cite{aa}. It also appears in a preferential selection model
for network evolution \cite{lipo}.

Below we study the above model for $f=$const, and show that it leads
to a non-trivial leadership scenario. To get richer scenarios, we
shall introduce additional variables.

\subsection*{Credibility scores}

For each agent $i$ we now introduce its credibility score
$\sigma_i(t)\geq 0$, which is a definite feature of an agent at a
given moment of time \cite{aronson}. Credibility refers to the
judgments made by a message recipient concerning the believability of
a communicator. A more general definition of credibility (not employed
here) should account for its subjective aspect; a message source may
be thought highly credible by one perceiver and not at all credible by
another \cite{aronson}.

Credibility scores interact with $m_i$ and $\tau_{ij}$ by modulating
the function $f$ in (\ref{eq:35}):
\begin{eqnarray}
  \label{eq:355}
f=f[\sigma_j(t)-\beta \,\sigma_i(t)], \qquad 
\beta=0, 1,
\end{eqnarray}
where we assume for simplicity
\begin{eqnarray}
  \label{eq:110}
  f[x]=x \quad {\rm for} \quad  x>0; \qquad 
f[x]=0 \quad {\rm for} \quad  x\leq 0.
\end{eqnarray}
Thus for $\beta=1$, the agent $i$ reacts only on those with
credibility score higher than $\sigma_i$, whereas for $\beta=0$ every
agent $j$ can influence $i$ proportionally to its score $\sigma_j$.
For convenience, we restricted $\beta=0,1$ to two values. (Note that
if we define $f[x]$ to be a positive constant for $x\leq 0$, then the
situation without scores can be described via $\beta\to\infty$.)


Dynamics of $\sigma_i$ is determined by the number of agents that
follow $i$ and by the amount of attention those followers pay to the
messages of $i$:
\begin{eqnarray}
  \label{eq:400}
  \sigma_i(t+1)=(1-\xi_1)\sigma_i(t)+\xi_2\sum_{k=1}^N m_i(t)
  \tau_{ki}(t),
\end{eqnarray}
where $\xi_1$ and $\xi_2$ are constants: $1\geq \xi_1> 0$ quantifies
the credibility score loss (forgetting), while the term with $\xi_2>0$
means that every time the agent $i$ activates, its score increases
proportional to the weight $\tau_{ki}(t)$ of its influence on $k$. If
$i$ is not active, ${m}_{i}(t)=0$, its score decays.

Development of complex network theory motivated many models, where the
links and nodes are coupled
\cite{sutton_barto,e-y1,e-y2,ito,rohl,armen,garshin,jost,caldarelli,chin};
see \cite{gross} for an extensive review. Neurophysiological
motivation for studying such models comes from the synaptic plasticity
of neuronal connections that can change on various time scales
\cite{peto,spike}.

The ingredients in the evolution of the credibility score
(\ref{eq:400}) do resemble the notion of fitness, as introduced for
models of competing animals \cite{bona}. There the fitness determines
the probability of winning a competition. Similar ideas were employed
for modeling social diversity \cite{naim}; see \cite{castel} for a
review.

\subsection*{Initial conditions }
\label{initial_conditions}

All agents are equivalent initially:
\begin{eqnarray}
  \label{eq:8}
 \sigma_i(0)=0. 
\end{eqnarray}
The initial network structure is random [cf. (\ref{eq:23})]
\begin{eqnarray}
  \label{do}
\tau_{ij}(0)=n_{ij}\left/\sum_{k=1}^{N} n_{ik}\right., ~~~
n_{ij}\in [0,b], ~~~ n_{ii}=0,
\end{eqnarray}
where $n_{ij}$ are independent random variables homogeneously
distributed over the interval $[0,b]$. Now
\begin{eqnarray}
  \label{strength}
\phi_i\equiv  \sum_{k=1}^N \tau_{ki}(0),
\end{eqnarray}
measures the initial cumulative influence of $i$ (i.e. its initial
social capital and estimates the initial rate of score generation;
cf. (\ref{eq:400}). (Financial capital is money. Physical capital is
tools, machinery {\it etc}. Human capital is people. Social capital is
the relationship among persons. Human capital resides in people;
social capital resides in the relations among them \cite{encyclo}.)

For $m_i(0)$ we impose initial conditions, where some agents are
activated initially (by a news or discussion subject), i.e. $m_i(0)$
are independent random variables:
\begin{gather}
  \label{eq:3}
  {\rm Pr}[m(0)=1]=\gamma, ~~~\,   {\rm Pr}[m(0)=0]=1-\gamma,~~~\,
\gamma\leq 1/2.
\end{gather}

\subsection*{Thresholds of collective activity}

The only way the initial activity can be sustained is if the agents
stimulate each other (as it happens in a real discussion process).
Our numerical results show that with initial conditions (\ref{eq:3})
there exists a sustained activity regime. Specifically, there are two thresholds
$\Q^+$ and $\Q^-$, so that for $q\geq \Q^+$ the initial activity is
sustained indefinitely,
\begin{eqnarray}
  \label{eq:1}
\sum_{i=1}^N m_i(t)>0, \quad {\rm if} \quad q\geq \Q^+.   
\end{eqnarray}
If $q\leq \Q^-$ the initial activity decays in a finite time $t_0$
(normally few time-steps) 
\begin{eqnarray}
  \label{eq:4}
\sum_{i=1}^N m_i(t\geq t_0)=0, \quad {\rm if} \quad q\leq \Q^-.   
\end{eqnarray}
For $\Q^+>q> \Q^-$ the activity sustaining depends on the realization
of random initial conditions $m_i(0)$ and $\tau_{ij}(0)$: either it is
sustained indefinitely or it decays after few steps. Qualitatively,
the activity is sustained if sufficiently well-connected agents are
among the initially activated ones; see below. $\Q^+$ and $\Q^-$
depend on all the involved parameters, their numerical estimates are
given below in (\ref{est}). Note that $\Q^->1$, since for $q<1$ no
activity spreading is possible: even the maximal weight $\tau_{ik}=1$
cannot activate $i$; see (\ref{n2}, \ref{eq:90}).


Emergent networks (described below) depend mainly on parameters
$\alpha$ and $\beta$; see Table \ref{tab1}. Other parameters can be
important for supporting activity, i.e. $\Q^+$ and $\Q^-$ depend on
them, but are not crucial for determining the type of the emerging
network. So we fix for convenience [cf. (\ref{eq:3}, \ref{do},
\ref{eq:400})]:
\begin{eqnarray}
  \label{eq:6}
  N=100, ~~ \gamma=0.5, ~~ b=10, ~~\xi_1=0.1, ~~\xi_2=1.
\end{eqnarray}
For these parameters and (for example) $\alpha=\beta=1$, we got
numerically
\begin{eqnarray}
  \label{est}
  \Q^-=2.19, \qquad \Q^+=2.65.
\end{eqnarray}

\subsection*{Behavioral noise}
\label{noise_noise}

The deterministic firing rule (\ref{n1}) can be modified to account
for agents with behavioral noise. We want to make it possible for an
agent to activate (not to activate) even for sub-threshold
(super-threshold) values of the potential.  The noise will be
implemented by assuming that the threshold $u_i+v_i(t)$ has (besides
the deterministic component $u_i$ discussed in (\ref{n1})) a random
component $v_i(t)$. These quantities are independently distributed
over $t$ and $i$. Now $v_i(t)$ is a trichotomic random variable, which
takes values $v_i(t)=\pm V$ with probabilities $\frac{\eta}{2}$ each,
and $v_i(t)=0$ (no noise) with probability $1-\eta$. Hence $\eta$
describes the magnitude of the noise. We assume that $V$ is a large
number, so that with probability $\eta$, the agent ignores $w_i$ and
activates (or does not activate) randomly. Thus, instead of
(\ref{n1}), we now have
\begin{eqnarray}
  \label{noise}
&&m_i(t)=\vartheta [\,\phi_i(t)(\, w_i(t)-u_i\,)\,],\\
  \label{n33}
  &&  {\rm Pr}[\phi_i(t)=1]=1-\eta, \qquad   {\rm Pr}[\phi_i(t)=-1]=\eta,~~
\end{eqnarray}
where $\phi_i(t)$ are independent (over $i$ and over $t$) random
variables that equal $\pm 1$ with probabilities ${\rm Pr}[\phi_i=\pm
1]$.

Qualitatively the same predictions are obtained under a more
traditional (for the neuronal network literature \cite{peto}) model
of noise, where the step function in (\ref{n1}) is replaced by a
sigmoid function.

\section*{\large Laissez-faire leadership}
\label{nocredit}

\begin{figure}
\begin{center}
{\bf \large Figures}
\end{center}
\centering
\includegraphics[width=0.4\columnwidth]{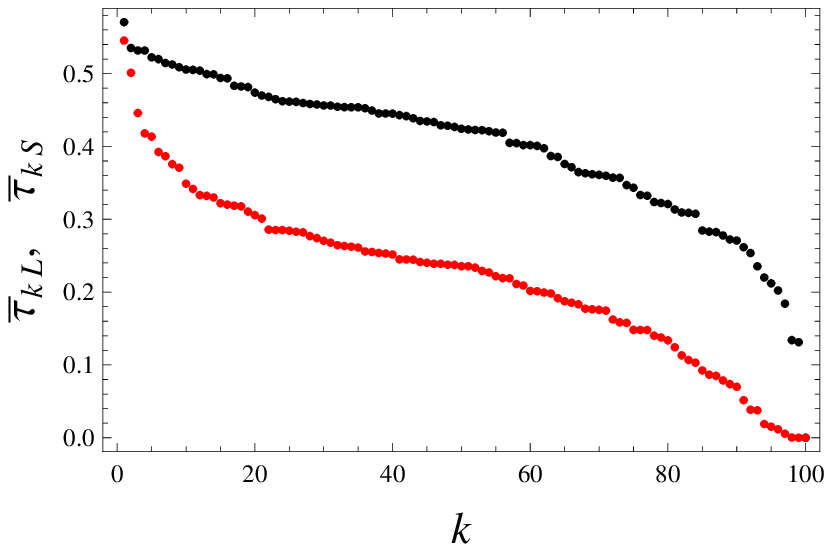} 
\caption{Laissez-faire leadership.\\ In (\ref{eq:35}) we set $f=1$ and
  $\alpha=1$. The behavioral noise is weak: $\eta=0.07$;
  cf. (\ref{noise}, \ref{n33}). The intergent coupling $q=2$ is
  sizable, but is smaller than (\ref{est}). The observation time:
  $T=600$; see (\ref{eq:7}). \\
  Black points (upper curve): the average weights $\bar{\tau
  }_{kL}$ by which the leader $\L$ influences other agents; see
  (\ref{eq:7}). Red points (lower curve): $\bar{\tau }_{kS}$ that
  quantify the influence of the leading sub-leader $\S$ on other
  agents. Here $\bar{\tau }_{kS}$ and $\bar{\tau }_{kL}$ were
  separately arranged in the decreasing order over $k$. }
\label{fig1}
\end{figure}

\begin{figure}
 \includegraphics[width=0.4\columnwidth]{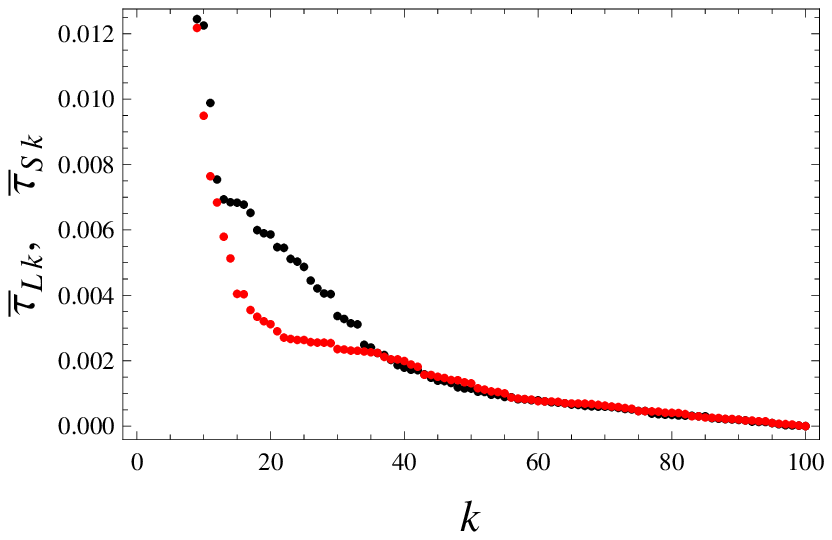} 
  \caption{Laissez-faire leadership.\\ The same parameters as in
    Fig~\ref{fig1}, but for the weights $\bar{\tau }_{Lk}$ and
    $\bar{\tau }_{Sk}$ that quantify the influence of followers on the
    leader $\L$, and on the sub-leader $\S$, respectively. Again,
    $\bar{\tau }_{Sk}$ and $\bar{\tau }_{Lk}$ were separately arranged
    in decreasing order.\\ Note that the influnce of $\L$ and on $\S$,
    and $\S$ on $\L$ are comparable: $\bar{\tau }_{SL}=0.545378$ and
    $\bar{\tau }_{LS}=0.535170$. }
\label{fig2}
\end{figure}

\begin{figure*}
\centering
\includegraphics[width=0.4\columnwidth]{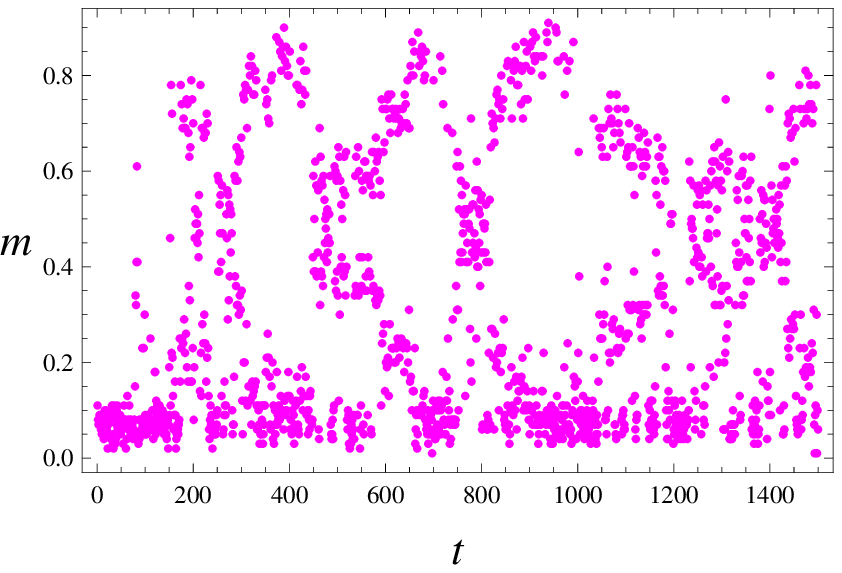}
\caption{Laissez-faire leadership.\\ The same parameters as in Fig~\ref{fig1}.  
Collective activity $m(t)=\sum_{i=1}^N m_i(t)$ versus time $t$. It is
seen that $m(t)$ displays irregular (noisy) behavior.}
\label{fig3}
\end{figure*}

\begin{figure*}
\centering
 \includegraphics[width=0.4\columnwidth]{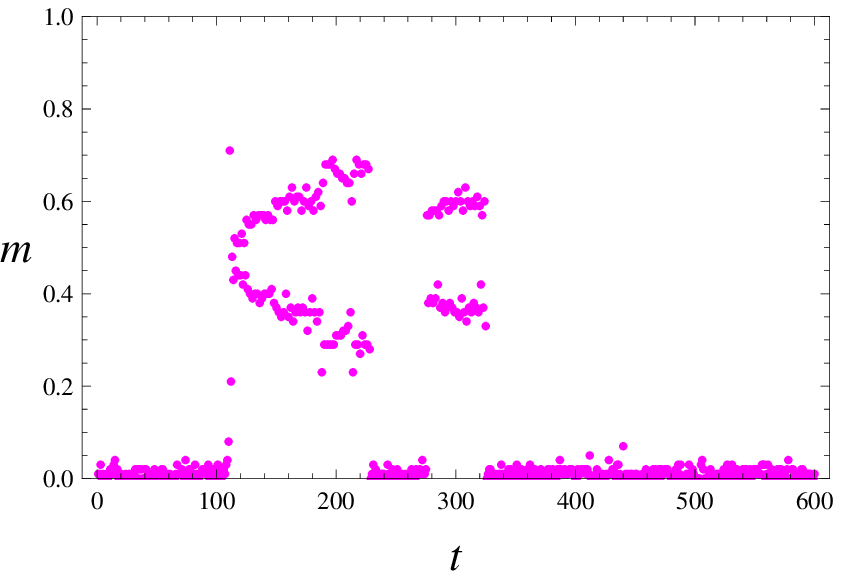}
\caption{Laissez-faire leadership. \\ The same parameters as in
  Fig~\ref{fig1}.  and Fig~\ref{fig3}.  Collective activity
  $m(t)=\sum_{i=1}^N m_i(t)$ versus time $t$, but for a weaker
  noise with magnitude $\eta=0.01$.  }
\label{fig4}
\end{figure*}

We shall study (\ref{n1}, \ref{n2}) under a weak-noise ($\eta\ll 1$ in
(\ref{noise}, \ref{n33})), but without scores, i.e.  we take
$f[x]=$const in (\ref{eq:35}). The magnitude $q$ of the inter-agent
interaction (see (\ref{eq:90})) is taken so that no activity is
sustained without the noise, i.e. $q<{\cal Q}^-$; cf.~(\ref{eq:4},
\ref{est}). Hence we are looking for a regime, where both noise and
inter-agent interactions are essential; see
Figs~\ref{fig1}--\ref{fig4}. The activity of the system|as measured
by $m(t)=\frac{1}{N}\sum_{i=1}^N m_i(t)$|is now much larger than the
noise magnitude $\eta$, i.e. the noise is amplified; see
Figs~\ref{fig3}. We shall see that in this regime there does emerge
a laissez-faire leadership scenario.

In this noisy situation, the network structure is studied via
time-averaged weights:
\begin{eqnarray}
  \label{eq:7}
\overline{\tau}_{ij}=\frac{1}{T}\sum_{t=s}^{T+s}\tau_{ij}(t),
\end{eqnarray}
where the observation time $T$ is sufficiently large.

The average, cumulative influence of an agent $k$ to all other agents
is quantified by $\sum_{i=1}^n\bar{\tau}_{ik}$. The leader can be
defined by the maximum of this quantity over $k$. We saw that there
emerges a leader ($\L$) that collects feedback from its many other
agents (followers); see Figs~\ref{fig1} and \ref{fig2}. There is
also a sub-leader $\S$ (or few sub-leaders depending on the
realization of the noise and the initial state) for which
$\sum_{i=1}^n\bar{\tau}_{ik}$ is next to maximal over $k$.  The
sub-leader $\S$ has its own followers and does collect feedback from
them. The influence of $L$ and $S$ on other agents is larger than the
back-influence of those agents; see Figs~\ref{fig1} and \ref{fig2}.
Many followers are shared between $\L$ and $\S$. All followers
influence each other, but with much smaller weights ${\cal
  O}(\frac{1}{N})$.

The relation between $\L$ and $\S$ is not hierarchical, since they
drive each other with comparable magnitudes; see Figs~\ref{fig1} and
\ref{fig2}. They are distinguished from each other by the fact that
$\L$ has more followers and influences them stronger.

The collective activity $m(t)=\frac{1}{N}\sum_{i=1}^N m_i(t)$ shows
irregular (chaotic) oscillations; see Fig~\ref{fig3}. Generally, the
emergence of $\L$ and $\S$ takes few such oscillations, i.e.  a rather
long time ($t\sim 500$ for parameters of
Figs~\ref{fig1}--\ref{fig3}). For even smaller magnitude of noise,
the system activates via relatively rare bursts; see Fig~\ref{fig4}.

This leadership is not an epiphenomenon. Indeed, we can intervene into
the system and block the activity of $\L$ and $\S$ by suddenly raising
their activation thresholds from $u=1$ to some very large value. As
compared to the same situation, but without intervention, the
time-averaged overall activity
\begin{eqnarray}
  \label{eq:9}
\frac{1}{TN}\sum_{t=s}^{T+s}\sum_{i=1}^N m_i(t)  
\end{eqnarray}
does decrease by $20-30\%$, and is recovered only after a long time,
when new leader and sub-leader emerge. (No statistically significant
activity reduction was seen when the same amount of non-leaders was
blocked). For the situation shown in Figs~\ref{fig1}--\ref{fig4},
the intervention was realized by letting the system to evolve for
$t=1,...,600$, suddenly blocking $\L$ and $\S$, taking $s=T=600$ and
then looking at the activity change (reduction) by calculating
(\ref{eq:9}) with and without intervention.

Note that the scenario is only weakly dependent on the value of
$\alpha$, but it is essentially based on the noise. For the noise-free
situation $\eta=0$ [see (\ref{noise}, \ref{n33})], we get either no
activity whatsoever (for a sufficiently small $q$), or a no-leader
activity-sustaining for a larger $q$. Another essential aspect is the
sufficiently fast adaptation, as quantified by the value of
$f[x]=$const in (\ref{eq:35}). If for parameters of
Figs~\ref{fig1}--\ref{fig4} we take $f=0.25$ (instead of $f=1$), no
activity above the noise magnitude $\eta$ is detected.

This scenario is similar to the laissez-faire leadership (as discussed
e.g. in \cite{encyclo,winkler,hogg}): substantially autonomous
followers, the existence of sub-leaders, no hierarchy between the
leader and sub-leader(s), irregular activity structure. And yet our
model shows that this is a real leadership: in contrast to what people
sometimes think and say about laissez-faire leaders, our model shows
that when blocking the leader and sub-leader the activity of the
system does decrease. Moreover, it takes a long time to establish a
laissez-faire leader and to replace it. We stress however that the
effect of the laissez-faire leadership is visible for sufficiently
long times and hence requires time-averaged indicators.

Thus the present model shows that the laissez-faire leadership
scenario emerges in a noisy (i.e. information rich) situation,
where the inter-agent interaction is sizable, but it not so strong
that the activity is sustained without noise.

\section*{\large Participative leadership}
\label{participative}

\begin{table*}[ht]
  \caption{Leadership scenarios for different parameters of
    (\ref{eq:355}) and $f[x]$ given by (\ref{eq:110}). For
    intermediate values of the parameters (e.g. $1<\alpha<1.5$) we get
  mixtures of the corresponding scenarios, or one of the scenarios is
  selected depending on initial conditions. }
\begin{tabular}{| c|c |c |c|}
\hline
 & $\alpha\gtrsim 1.5$  & $\alpha\simeq 1$ & $\alpha\lesssim 0.5$ \\
\hline
$\beta=1$ & hierarchy       & single leader  & succession of leaders\\
          & (participative) & (participative)  &  (autocratic) \\ 
          &                 &                   &               \\
\hline
$\beta=0$ & hidden leaders  & leader + helper  & succession of leaders \\
          & (autocratic)    & (autocratic)     & (autocratic)\\ 
\hline
\end{tabular}
\label{tab2}
\end{table*}

\subsection*{Single  participative leader}
\label{single_demo}

\begin{figure*}
\centering
 \includegraphics[width=0.5\columnwidth]{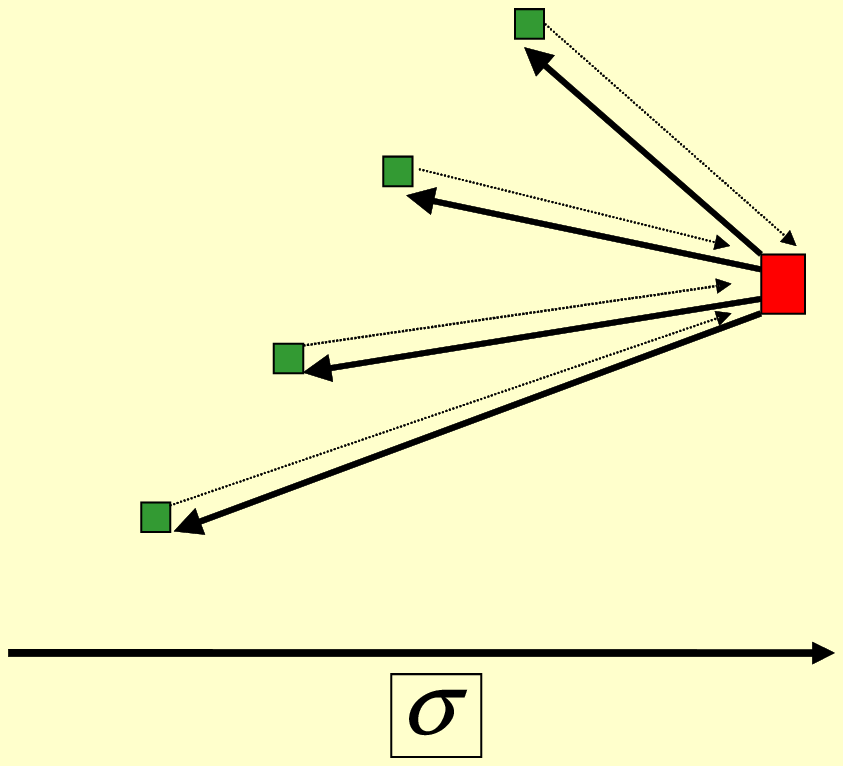} 
\caption{Participative leadership.\\ Emergent network structures
  according to Eqs~(\ref{n1}--\ref{do}, \ref{eq:3}). Parameters
  are chosen from (\ref{eq:6}, \ref{est}) and $q>\Q^-$. \\
  {\it Single Participative leader}: $\alpha=\beta=1$; see
  (\ref{eq:35}, \ref{eq:355}). The leader $\L$ (red square) has the
  highest score ($\simeq 500$) and stimulates all other agents
  (followers, green squares) with the maximal weight $\tau_{i{\rm
      L}}=1$ (bold arrows). Followers (green squares) have different
  credibilities $\sigma_i={\cal O}( 1/N)$; each of them stimulates the
  leader with weights $\tau_{Li}={\cal O}( 1/N)$: a follower with a
  larger score influences the reader stronger. }
     \label{fig5}
\end{figure*}

We solve dynamics with credibility scores (\ref{n1}, \ref{n2},
\ref{eq:35}, \ref{eq:355}, \ref{eq:110}, \ref{eq:400}) for $q>\Q^-$;
hence there is a possibility for the sustaining activity.  Noise is
omitted, $\eta=0$ in (\ref{noise}, \ref{n33}), because it is
irrelevant; see below.  The parameters in (\ref{eq:35}, \ref{eq:355})
are
\begin{eqnarray}
  \label{eq:5}
\alpha=\beta=1.   
\end{eqnarray}
After $\simeq 20$ time-steps the system enters into a stationary
state, where the activation frequencies do not depend on time.  There
emerges a leader ($\L$) that drives all the other agents (followers)
with the maximal weight $\tau_{k\L}=1$ [cf. (\ref{n1}, \ref{n2})],
i.e. each follower is influenced only by $\L$; see Fig~\ref{fig5} for
schematic representation. Hence $\L$ emergence in a 
much shorter time than the laissez-faire leader, and $\L$ also
influences the followers much strongly. 

Credibility scores $\sigma_k$ of followers and their influence weights
$\tau_{\L k}$ reach quasi-stationary values over a larger time
[$\simeq 100$ for parameters of (\ref{eq:6})]. $\L$ has the largest
final score ($\sigma_{\L}\gg 1$). Followers have much lower scores and
influence $\L$ via smaller weights $\tau_{\L k}={\cal
  O}(1/N)<1$. Followers do not influence each other (in contrast to
the laissez-faire scenario) precisely because the influence of $\L$ on
its follower is maximal; see (\ref{eq:900}). The activity of the
leader (hence the overall activity) is sustained due to cumulative
feedback from followers to $\L$.  Hence this is a participative
leadership scenario.

\begin{figure*}
\centering
 \includegraphics[width=0.5\columnwidth]{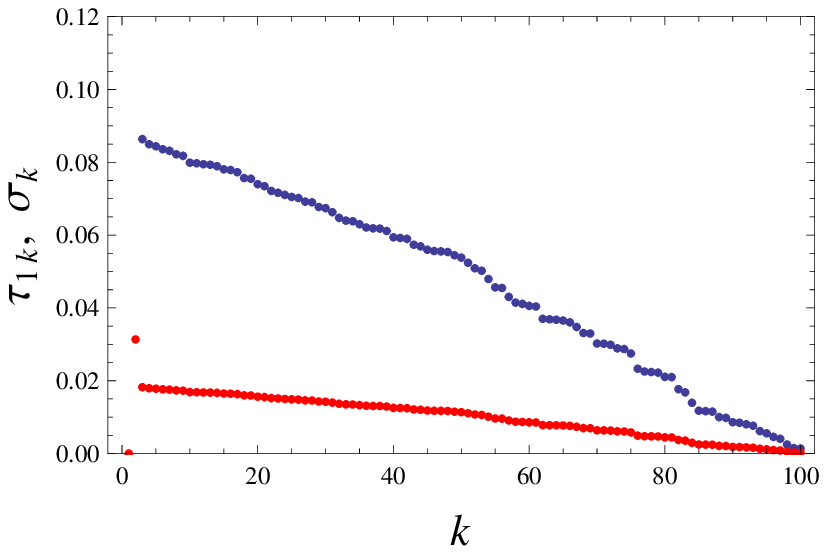} 
\caption{Participative leadership: single emergent leader. \\
  Distribution of stationary credibilities $\sigma_k$ (blue points,
  upper curve) and weights $\tau_{1k}$ (red points, lower curve) by
  which the agent with rank $k$ ($N\geq k\geq 2$) influences the
  leader ($k=1$).  The agents are ranked according to their final
  score: $k=1$ is the highest-score agent (leader), $k=N$ is the
  lowest score agent.  Eqs~(\ref{n1}--\ref{eq:400}, \ref{eq:3}) are
  solved for (\ref{eq:6}) and $q=2.5$. The dynamics was followed by
  $200$ time-units. }
\label{fig6}
\end{figure*}

\begin{figure*}
\centering
 \includegraphics[width=0.5\columnwidth]{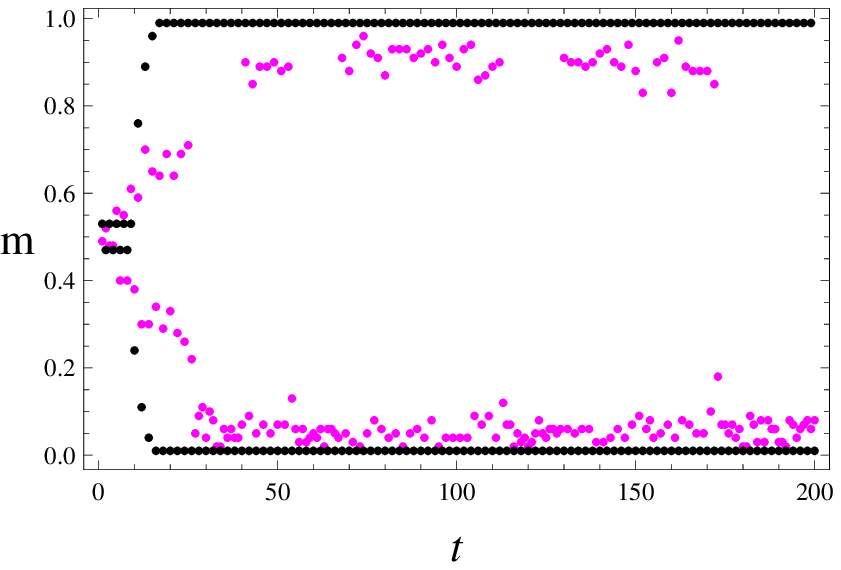} 
\caption{Participative leadership. Single emergent leader. \\
  The collective activity $m(t)=\frac{1}{N}\sum_{k=1}^N m_k(t)$ as a
  function of time $t$ for $q=2.7$ (other parameters are the same as
  in Fig~\ref{fig6}). Black points (two straight lines): the noiseless
  situation $\eta=0$. Magenta points: $\eta=0.05$; cf. (\ref{noise},
  \ref{n33}). In the noiseless situation $m(t)$ takes only two values
  $0.01$ (the leader is active) and $0.99$ (followers are active). For
  the noisy situation $m(t)$ assumes two well-separated sets of values
  at $\sim 0.1$ and $\sim 0.9$, respectively.  }
\label{fig7}
\end{figure*}

The score distribution of followers is such that a follower with a
higher score influences the leader more; see Fig~\ref{fig6}. In the
noiseless situation the dynamics is strictly synchronized: only $\L$
fires in one time-unit. In the next time-unit $\L$ is passive, while
followers activate together. The strict synchronization disappears
after introducing a small noise; see (\ref{noise}, \ref{n33}) and
Fig~\ref{fig7}, where the collective activity
$m(t)=\frac{1}{N}\sum_{k=1}^N m_k(t)$ is monitored with and without
noise. Other characteristics of the scenario stay intact.

Initially, $\L$ had to be among activated agents: $m_{L}(0)=1$, {\it
  and} it was most probably having the largest social capital [in the
sense of (\ref{strength})] among the initially activated agents; this
is the case in $\simeq 95 \%$ of (random) initial conditions. Hence
the leader emerges due to the amplification of its initially small
advantages over other agents.

Comparing the laissez-faire leader with the present one, we see that
introducing credibility scores enforces a hierarchy, i.e. it
eliminates connections between the followers and maximizes the
influence of the leader.

\subsection*{Externally imposed versus emergent leadership}

It is of clear importance to understand when and whether externally
imposed leaders can compete with those that emerged from within the
group \cite{encyclo}. We model imposed leaders by externally driving
(sponsoring) the activity of certain agent(s). To this end, we add a
term $(1-m_i(t)\,)r_i$ to the right-hand-side of (\ref{n2}), where
$r_i\geq 0$ is the rate of potential generation that does not depend
on inter-agent interaction. In particular, $r_i$ can be an externally
imposed rate.  Thus (\ref{n2}) becomes
\begin{gather}
  \label{eq:10}
  w_i(t+1)=(1-m_i(t)\,) \sum_{j=1}^N q_{ij}(t)m_j(t)+(1-m_i(t))r_i . 
\end{gather}
We focus on the situation, where only one agent is externally driven:
\begin{eqnarray}
  \label{eq:2}
r_1>1, \qquad  r_{i\geq 2}=0,
\end{eqnarray}
i.e. the first agent activates with the maximal frequency $0.5$;
cf. (\ref{n1}, \ref{n2}).

Now for $q<{\cal Q}^-$ the first agent becomes a leader. For ${\cal
  Q}^-<q<{\cal Q}^+$ (and (\ref{eq:5})) it can|depending on initial
conditions|become a participative leader, and it generally does not
become a leader for $q>{\cal Q}^+$: the emergent leader takes over the
externally driven agent.

\subsection*{Hierarchy of leaders}

\begin{figure*}
\centering
 \includegraphics[width=0.5\columnwidth]{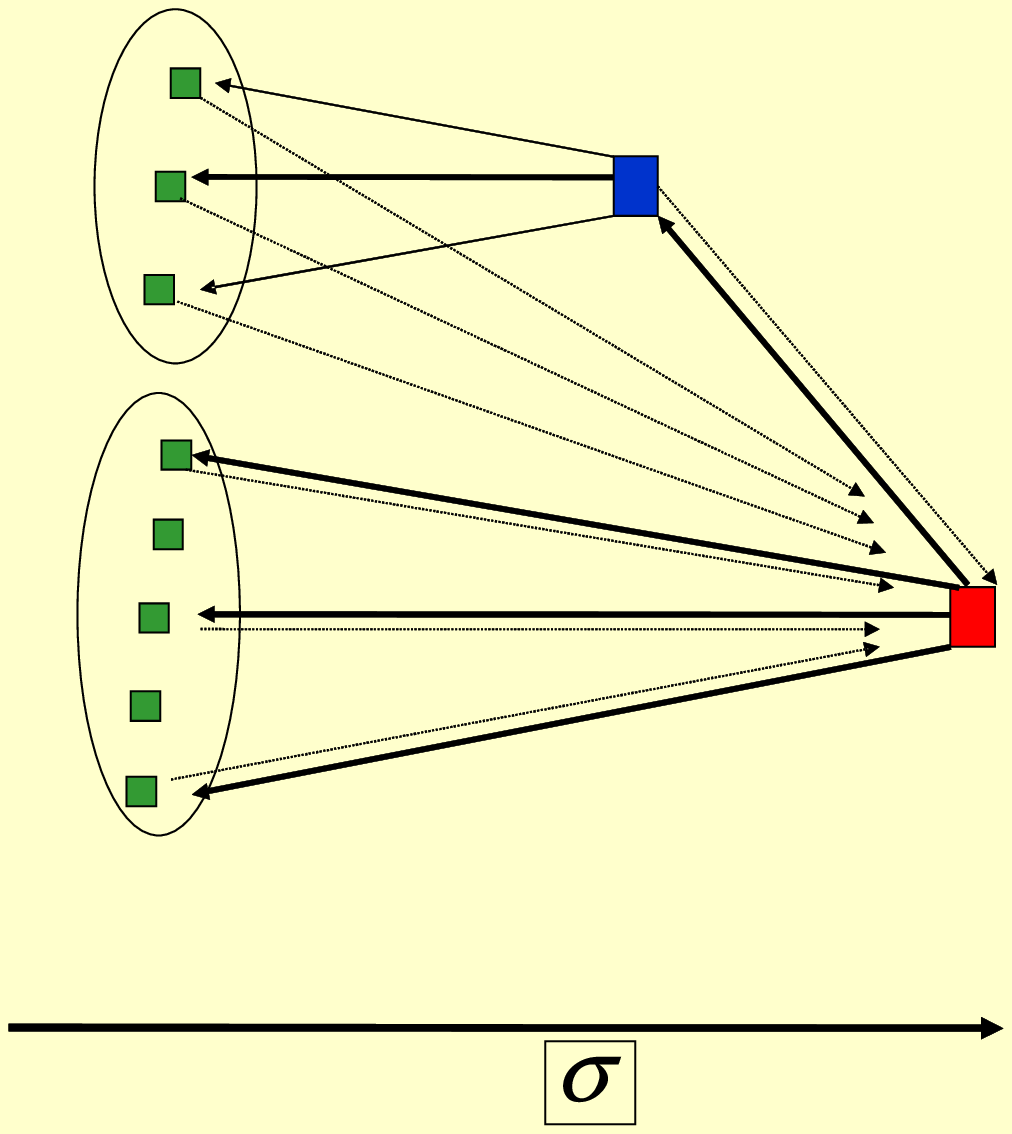} 
\caption{Participative leadership.\\
  {\it Hierarchic leadership}: $\alpha=2$, $\beta=1$. (Other
  parameters are the same as in Fig~\ref{fig5}.) The followers are
  divided into two groups, strongly driven by respectively leader
  (red) and sub-leader (blue). The feedback is collected by the leader
  only. }
     \label{fig8}
\end{figure*}

If instead of (\ref{eq:5}) we employ $1.5 \lesssim\alpha$ and
$\beta=1$, we get a hierarchic leadership scenario; see
Fig~\ref{fig8}. There emerge one leader and few sub-leaders. The
precise number of sub-leaders depends on initial conditions.  Each
sub-leader has followers that are not directly influenced by the
leader. The latter influences only their sub-leader; see
Fig~\ref{fig8}. In a sense the leader delegates some of its
influence to the sub-leader(s).  But the main feature of the
participative leadership is kept: only the top leader collects
feedback from all other agents.

Note the difference with sub-leaders within the laissez-faire
scenario: there $\bar{\tau}_{\L \S}\simeq\bar{\tau}_{\S\L}$, i.e. the
leader and sub-leader influence each other with comparable weights. In
contrast here $\tau_{\L\S}\ll \tau_{\S\L}=1$, i.e. the influence of
the leader is maximal and is much larger than the influence of the
sub-leader on the leader. Thus the participative leadership is
hierarchical, in contrast to the laissez-faire scenario. This
hierarchy is introduced by credibility scores that were absent in the
latter scenario.



\section*{\large Autocratic leadership}
\label{autocratic}

\subsection*{Single autocratic leader}

We turn to studying (\ref{n1}, \ref{n2}, \ref{eq:35}, \ref{eq:355},
\ref{eq:110}, \ref{eq:400}) with
\begin{eqnarray}
  \label{eq:505}
\alpha=1, \qquad \beta=0,   
\end{eqnarray}
and for $q>\Q^-$, i.e., for initial conditions where the activity can
be sustained. We get ${\cal Q}^-=2.18$ under (\ref{eq:505},
\ref{eq:6}); cf. (\ref{est}).

\begin{figure*}
\centering
 \includegraphics[width=0.5\columnwidth]{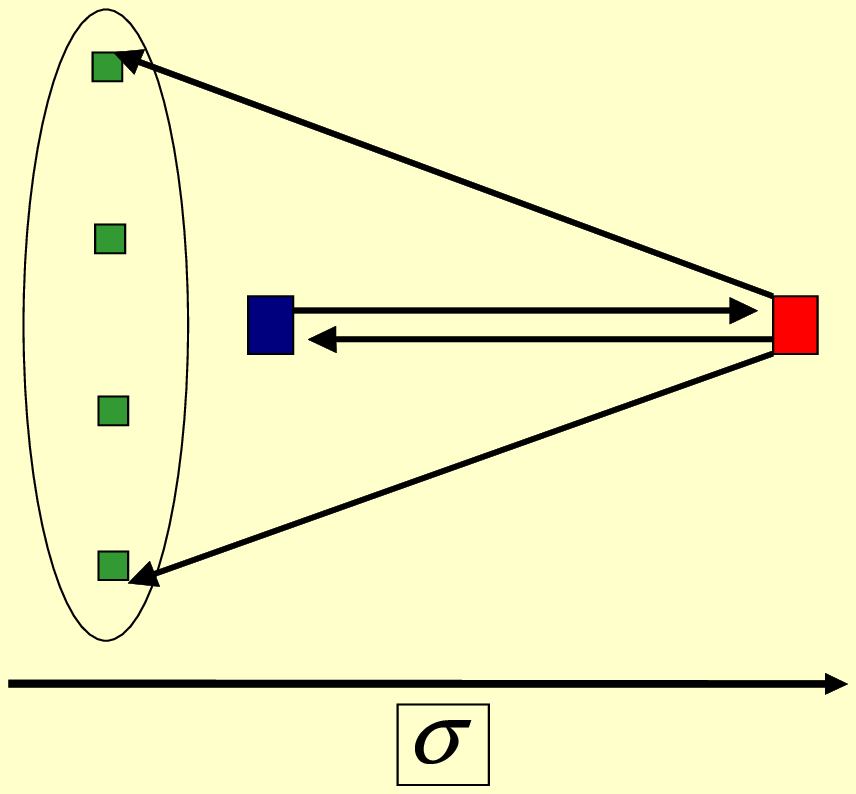}
    \caption{Autocratic leadership.\\
      {\it Single autocratic leader}: $\alpha=1$, $\beta=0$ and
      $q>\Q^-$; see (\ref{eq:35}, \ref{eq:355}, \ref{eq:1},
      \ref{eq:4}). Other parameters are chosen according to
      (\ref{eq:6}).  The leader (red square) stimulates others (green
      squares) and is stimulated by the helper (blue square). All
      these stimulations have the maximal weight equal to $1$. All
      other agents are passive spectators with zero score.  }
\label{fig9}
\end{figure*}

\begin{figure*}
\centering
 \includegraphics[width=0.5\columnwidth]{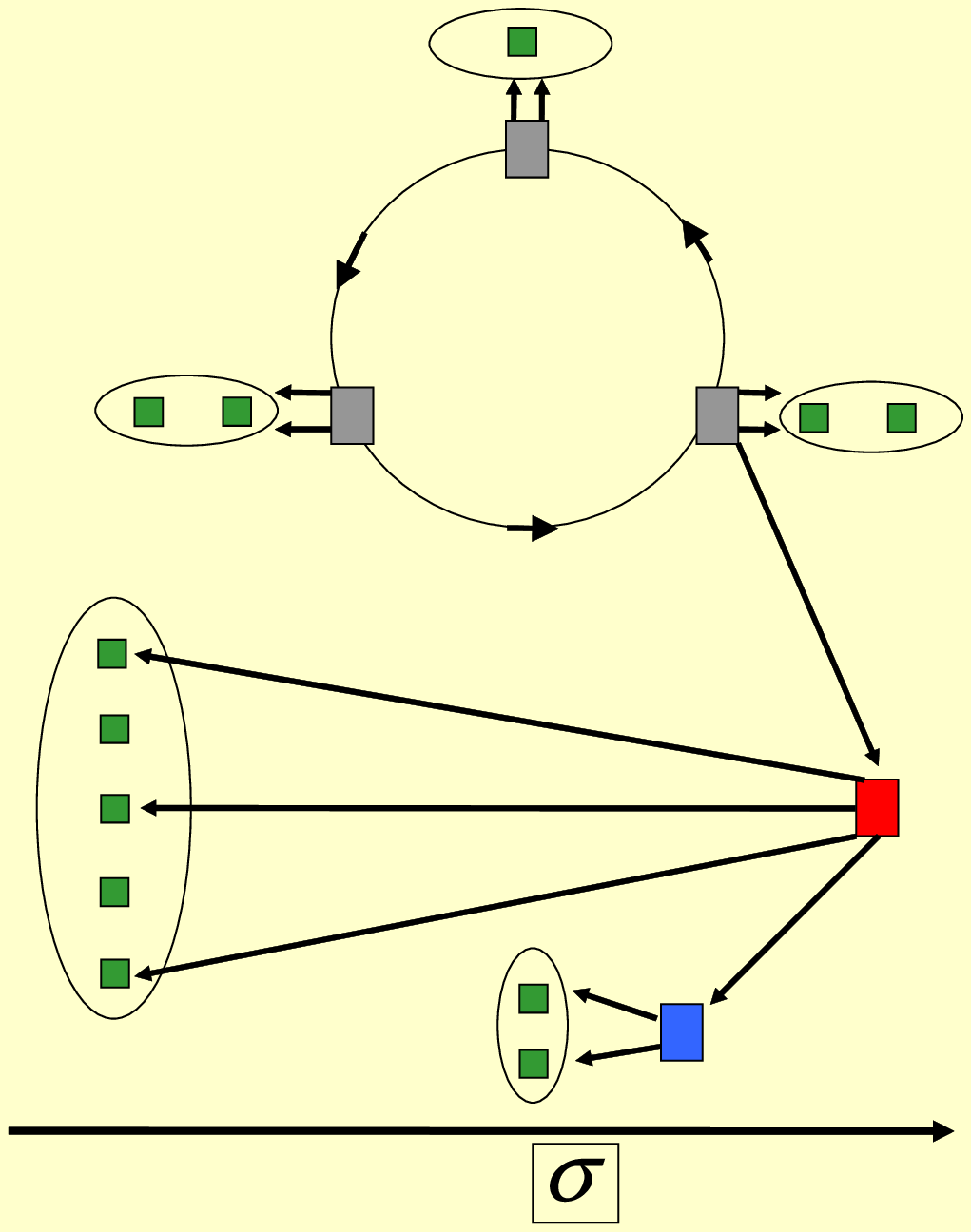}
    \caption{Autocratic leadership.\\
      {\it Hidden leadership}: $\alpha=1.5$, $\beta=0$ (other
      parameters are those of Fig~\ref{fig9}). The highest-score
      agent (red square) is driven by a circle of agents that drive
      each other cyclically (grey squares). The official leader (red
      square) has the largest number of followers (green squares),
      though each gray agent can have its own followers. All followers
      are driven by the maximal weight, have neglegible credibilities
      and do not feedback. }
     \label{fig10}
\end{figure*}

In $\simeq 20$ time-steps, there emerges a leader $L$ that has the
highest score and that influences all other agents with the maximal
weight $\tau_{kL}=1$; see Figs~\ref{fig9}, \ref{fig10}. However, it
gets feedback only from one agent $H$ that emerges together with $L$,
and that stimulates $L$ with the maximal weight, $\tau_{\rm
  LH}=1$. Hence, $H$ connects to no other agent. The only role of $H$
is that it stimulates the leader; see Figs~\ref{fig9},
\ref{fig10}. The long-time score of $H$ is next to the largest, e.g.,
for $q=2.5$ and (\ref{eq:6}) we get $\sigma_L\simeq 500$,
$\sigma_H\simeq 5$, while $\sigma_{k}\simeq 0$ for $k\not=L,H$.

For all other agents $k$ ($k\not =L,H$), the influence on $L$, on $H$,
and on each other (i.e., the magnitude of $\tau_{ik}$ for $k\not=L$ and
$k\not=H$) is negligible; they are completely passive followers. We
have here an autocratic leadership scenario, since no feedback is
present from the followers to the leader; see Figs~\ref{fig9},
\ref{fig10}.

A pertinent question is whether the autocratic $L$ had to have
(initially) a large social capital; cf. (\ref{strength}). The answer
is definitely no.  In $\simeq 50\%$ of initial conditions $L$ did not
have a large social capital initially, although both $L$ and
$H$ have to be among the initially active agents $m_L(0)=m_H(0)=1$;
cf. (\ref{eq:3}, \ref{eq:6}). In contrast to the participative
scenario, the selection of the autocratic leader is not fully
determined by its social capital.

Another difference with respect to the participative scenario is that
now $Q^+=\infty$. For whatever the large value of $q$, there are
initial states where the activity is not sustained; in particular, it
ceases before any leader emerges.  In this sense the autocratic
leadership is less stable. (However, the scenario is stable with
respect to introducing the noise (\ref{noise}, \ref{n33})). To
illustrate this point, note that the activity is not sustained|and
hence no definite network structure emerges|for regular initial
conditions $\tau_{ij}(0)=1/(n-1)$ (instead of (\ref{do})). For the
participative situation this homogeneous initial network structure
only delayed (by an order of magnitude) the convergence towards the
stationary structure. Thus, the emergence of the autocratic leader
(and its helper) demands initial inhomogeneity of the network.

Another implication of $Q^+=\infty$ is that the externally driven
agent (cf. (\ref{eq:2})) does always have a chance to become an
autocratic leader.

\subsection*{Hidden leadership}

We continue to focus on $\beta=0$, but now $1.5\lesssim\alpha$;
cf. Table \ref{tab2}. Instead of a single helper $H$, we get
(depending on the initial conditions) a few agents $H_1,H_2,...$; see
Fig~\ref{fig10}. The agents act on each other {\it cyclically}, and
although they have lower credibility scores than $L$, the one with the
highest score ($H_1$) drives $L$ with the maximal weight
$\tau_{LH_1}=1$. Thus $H_1,H_2,...$ are {\it hidden and real}
leaders. They are hidden because their score is not large; it is
smaller than the score of $L$, but it is real because they influence
each other strongly, and one of them influences $L$, which has the
maximal score.  Importantly, the influence of $H_1$ on $L$ is one-way;
$H_1$ does not get any feedback from $L$, in contrast to the previous
scenario with the single helper. Thus, within the group $H_1,H_2,...$
the score is not important, since these agents drive each other
cyclically. However, $L$ is still driven by the highest-score (among
$H_1,H_2,...$), agent $H_1$.

Three further details can be present in this scenario, depending on
initial conditions. First, it is also possible that $L$ belongs to the
group of hidden leaders $H_1,H_2,...$.  Second, the scenario may be
accompanied by a hierarchic structure, where the leader influences a
sub-leader that drives its own group of followers; see
Fig~\ref{fig10}. Third, some of $H_k$ (most frequently $H_1$) can
have their own followers, that are not the followers of $L$. Still,
$L$ has the largest number of followers and thereby the largest
credibility score; see Fig~\ref{fig10}.


\subsection*{Coalition of autocratic leaders (duumvirate)}
\label{coalition}

\begin{figure*}
  \centering
 \includegraphics[width=0.5\columnwidth]{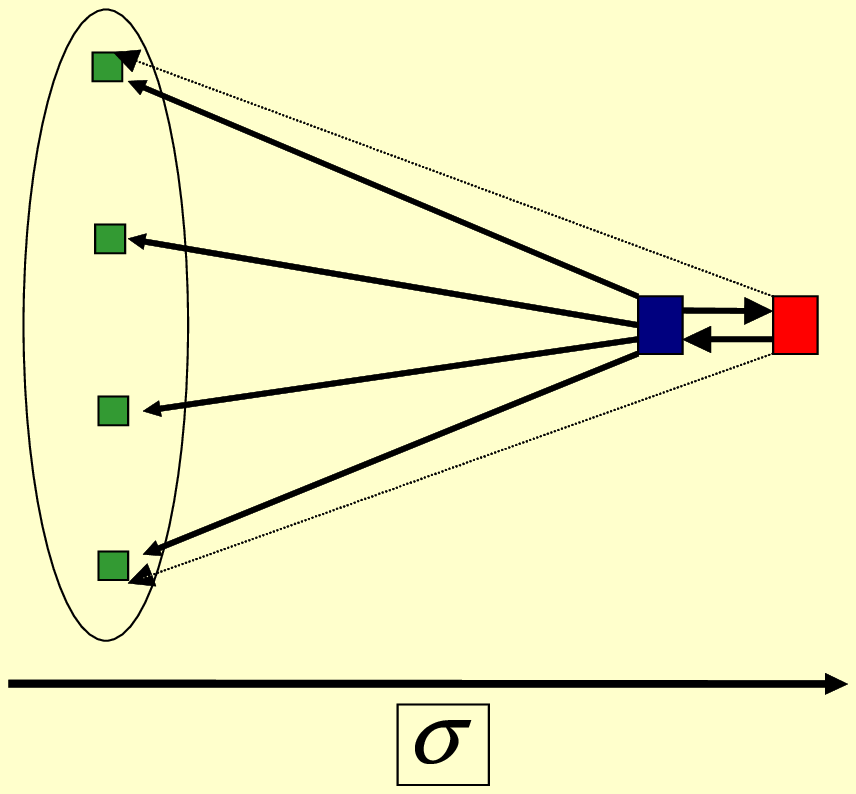} 
\caption{{\it Coalition} of two leaders.\\ Here $\alpha=0.25$,
  $\beta=1$ and (\ref{eq:6}). The red (blue) agent has the highest
  (one but highest) score. However, the real leader is now the blue
  agent, since it stimulates all other agents with the weight equal to
  $1$. The red agent stimulates the blue one with the weight $1$, and
  all other agents with the weight $\simeq 1/N$. All green agents are
  passive spectators with score close to zero. }
     \label{fig11}
\end{figure*}

If for $\beta=1$, $\alpha$ decreases from $\alpha=1$, then around
$\alpha=0.75$ (for parameters of (\ref{eq:6})) there is a change in
the final network structure. For $\alpha<0.75$ this structure is such
that there emerge two leaders ($L_1$ and $L_2$), whose scores are
approximately equal and much larger than the scores of all other
agents; see Fig~\ref{fig11}. They strongly drive each other (in the
final state): $\tau_{L_1L_2}=\tau_{L_2L_1} =1$). They drive {\it by
  turns} all other agents $k$; if $\tau_{kL_1}(t)\simeq 1$ and
$\tau_{k L_2}(t) = {\cal O}(1/N)$, then after a few time-steps $\delta$,
$L_1$ and $L_2$ interchange their roles: $\tau_{kL_2}(t+\delta)\simeq
1$ and $\tau_{kL_1}(t+\delta)= {\cal O}(1/N)$. For the noiseless
situation ($\eta=0$ in (\ref{noise}, \ref{n33})) we get $\delta=1,2$.
If a weak noise is present see (\ref{noise}, \ref{n33}), $\delta$
becomes a random number.

All
the other agents besides $L_1$ and $L_2$ are passive
spectators without influence on anyone.



The transition from a single autocratic leader to the successive
leadership takes place also for $\beta=0$, but for
$\alpha<0.6$ for parameters of (\ref{eq:6}).

If $\alpha$ is close to zero (e.g., $\alpha\leq 0.1$ for $\beta=1$ or
$\beta=0$), the relaxation to the stationary network structure is
slow. For times $t\sim 30-40$ there emerges a small (up to $10$
agents for the overall number of agents $100$) group of high-score
leaders that influence each other and drive the remaining majority of
agents without getting any sizable feedback from this majority. Only
on much longer times, $t\sim 1000-2000$, two leaders with the symmetric
interaction emerge from within this group. Other agents from the
high-score group move to the zero-score majority.  Thus, there can be a
larger group of leaders (more than two agents), but it is metastable.

\section*{\large Conclusion}

Leadership has been at the focus of many disciplines: history (many
biographies are about leaders), political science (governance
structure and function), philosophy (principles of good versus bad
leadership), management science (leadership practices), social
psychology (influence), communication research (leadership frequently
goes via|and is about|communication), and complex systems
\cite{encyclo,winkler}. But we lacked a tractable model that can
provide a theoretical laboratory for describing and studying
leadership scenarios.

We studied a formal, mathematical model for the emergence of opinion
leader in a collective of agents that are modeled via threshold
elements (neurons) living on a network. Agents can activate and
influence their neighbors via the network possibly making them active
as well. Weights of the network links are dynamic. They account for
the importance given by the agent to a specific link. The behavior
rules of the agents are summarized as follows; cf. Table \ref{tab1}.

-- The overall influence on an agent from its neighbors (i.e. its
attention) is limited. 

-- An agent re-distributes its attention via changing the link weights
such that more active agents get more attention. The conservatism of
the agent during this process is described by a parameter $\alpha\geq
0$; see (\ref{eq:35}). For $\alpha\to 0$ the agent tends to revise
even those links that were once deemed to be unimportant.

-- There is a possibility of involving into the model an agent's
credibility score, a dynamic variable that grows whenever the agent
actively influences its neighbors; otherwise the score decays. The
score couples to attention such that agents tend to be influences by
their higher-score neighbors. The extent to which they account for
their existing score is modeled by a parameter $\beta= 0,1$; see
(\ref{eq:355}). For $\beta=0$ an agent does not compare its own score
with the score of its neighbors.

-- Agents may be subject to noise, i.e. they activate or deactivate
randomly.

For this model we uncovered several types of leadership|depending on
$\alpha$, $\beta$, the noise, and the importance of credibility scores
(see Table \ref{tab2})|that do correspond to basic scenarios known
from social psychology and from real life.

$\bullet$ The laissez-faire leader emerges under weakly-noisy dynamics
without scores. Here, the overall activity is irregular, followers are
influenced weakly, their mutual interaction is not suppressed, and
autonomous sub-leaders with their followers are allowed. Some
followers are shared between the leader and sub-leader. The leader and
sub-leader(s) influence each other with comparable weights. The
emergence of the laissez-faire leader takes a lengthy time, but it is
a real leadership, since the overall activity decays (though not
completely) after suppressing the leader. The replacement of the
laissez-faire leader occurs spontaneously and also takes a lengthy
time. An important aspect of the laissez-faire leadership is that its
presence and its effect is visible only for sufficiently long
times. We uncovered this scenario by looking at time-averages.

Other leadership scenarios emerge after introducing the scores. This
emergence takes much shorter time.

$\bullet$ A participative (democratic) leader does influence all
agents strongly, hence it suppresses inter-follower
interactions. Still this leader accepts feedback from the followers:
it is active precisely due this feedback. If sub-leaders are allowed,
they are delegates of the leader. The latter does not influence
followers of the sub-leader, but strongly drives the sub-leader, whose
back-influence on the leader is weak. The leader emerges due to its
social capital (well-connectedness in the network). Initially its
social capital is only slightly larger than for the others, but it is
amplified dynamically due to credibility scores, and finally leads
this agent to leadership.  Within this scenario, no leader can be
imposed externally, if the inter-agent coupling is sufficiently
strong.

$\bullet$ The autocratic leadership scenario is realized for $\beta=0$
(see Table \ref{tab2}), i.e. when the agents do not respect their
credibility scores when re-distributing their attention. This point is
somewhat unexpected, naively one would think that the neglect of one's
own score is a route to democracy. An autocratic leader (i.e., the
highest score leader with the most followers) does not accept feedback
from its followers; they are completely passive. Instead, the activity
of the leader is supported either by symmetric interaction with a
single agent (helper), or by one-way driving from a group of hidden
(i.e., lower-score) agents. Hence, these agents are the real
leaders. No simple criteria was found for predicting the autocratic
leader from the initial state. In a sense, this leader is selected for
random reasons---a point that is anticipated to be one of the main
dangers of autocratic organizations \cite{madow}. The autocratic
scenario is susceptible to perturbations of the initial state (in
contrast to the participative leadership); i.e. for certain initial
states the activity of the system ceases because no leader is
established. One aspect of this susceptibility is that an autocratic
leader may be always imposed externally, again in contrast to the
participative leader. For non-conservative agents ($\alpha\to 0$), the
model predicts a coalition of autocratic leaders. However, the
coalition is meta-stable; for long times it reduces to a pair of
autocratic leaders that symmetrically interact with each other, and by
turn drive the remaining agents.

The above scenarios were established under concrete dynamic rules for
the behavior of agents, but it is likely that they will emerge more
generally, i.e., for other rules. A detailed understanding of this
premise, as well as empiric validations of the model, are left for
future work. Here we only stress that there are several directions via
which such a validation can progress. Within the main direction, one
can check possible leadership scenarios in various examples of social
media. Here the main problem is that normally the real network of
influences between agents is not known. However, several methods were
developed recently to resolve this problem
\cite{ha,huber,gomez_leskovec,barba1,barba2}, and since all the ingredients of
our model (e.g. the attention restriction {\it etc}) have their
analogues for agents of social media, one can hope to find a
reasonable classification of social media leaders that does resemble
the one studied here (i.e. laissez-faire, participative, autocratic,
weak and strong forms of hierarchy and of influence sharing,
duumvirate {\it etc}). 

There are few other directions along which one can attempt to validate
the present model. Recently, there was an interesting discussion on
the emergence of leaders in collective of robots \cite{robots}. Once
it is understood from our model that rather complex leadership
scenarios can emerge from simple behavioral rules, one can model these
rules for robots and look at their emergent leadership behavior. Next,
there is a large body of work devoted to various situations of
decentralized control; see e.g. \cite{decentralized}. By their very
definition such scenarios imply the absence of any
leadership. However, our results hint that claims on the absence of
leadership may in fact be overstated, and that laissez-faire leaders
may be hidden in (at least) some scenarios of decentralized control.

\section*{\large Acknowledgements}

This research was supported by DARPA grant No. W911NF-12-1-0034. 

\comment{
\section*{\large Supporting Information}

{\bf S1 Data}

The file ${\rm laissez\_\,faire\_\,supplementary.nb}$ provides a
simple numerical code (written in Mathematica) for simulating one of
the leadership scenarios discussed in the main text.
}


\clearpage

\end{document}